# Absolute Properties of the Triple Star CF Tauri

Claud H. Sandberg Lacy[1,4], Guillermo Torres[2], and Antonio Claret[3]

[1]Physics Department, University of Arkansas, Fayetteville, AR 72701, USA;

clacy@uark.edu

[2]Harvard-Smithsonian Center for Astrophysics, 60 Garden Street, Cambridge, MA

02138; gtorres@cfa.harvard.edu

[3]Instituto de Astrofísica de Andalucía, CSIC, Apdo. Postal 3004, E-18080 Granada,

Spain: claret@iaa.es

# ABSTRACT

CF Tau is now known to be an eclipsing triple star with relatively deep total and annular eclipses. New light and radial velocity curves as well as new times of minima were obtained and used for further modeling of the system. Very accurate (better than 0.9%) masses and radii of the eclipsing pair are determined from analysis of the two new light curves, the radial velocity curve, and the times of minimum light. The mass and luminosity of the distant third component is accurately determined as well. Theoretical models of the detached, evolved eclipsing pair match the observed absolute properties of the stars at an age of about 4.3 Gy and *[Fe/H]* = -0.14.

1. Introduction

In stellar astronomy it is still important to test current theory against accurate observations of stellar dimensions, masses, luminosities, and internal structure constants. These fundamental data can be supplied by measurements of eclipsing binary stars. Light curves, radial velocity curves, and O-C diagrams can be built up from measurements of brightness, Doppler shifts, and times of minimum light. The projected rotation rates (v sin i) of the components also may be measured from high-resolution spectra. In systems with distant 3$^{rd}$ components, the light travel time across the wide orbit can be measured to determine accurately some properties of this third star as well. Measured absolute properties can be compared with the results of stellar evolution theory in order to gauge the completeness of that theory, which is the main motivation of this type of investigation. A general compilation and investigation of these types of results is given by Torres, Andersen, & Giménez (2010).

The eclipsing binary star CF Tau, which we find below to be a detached evolved binary, is a relatively bright star ($V$ = 10.35 mag, Høg et al. 2000), originally classified as G0 (HD Catalogue) but, as a result of our study, it is now known to be a G0+K0 eclipsing pair with a much fainter 3$^{rd}$ component. The eclipsing components will be denoted Aa (the more massive, cooler, and larger eclipsing star), Ab (the less massive, hotter, and smaller eclipsing star), and the distant third component will be refered to as star B. The system was first discovered as a variable star by Morgenroth (1934). The early history of its observations is given by Szafraniec (1960), who also measured a visual light curve, which shows a shallow secondary eclipse indicating a much cooler eclipsing component.

It is included in lists of chromospherically active stars (Strassmeier et al. 1993, Kashyap & Drake 1999, Eker et al. 2008) and stars with X-ray and radio emission (Drake et al. 1992). Popper (1996) noted that he could see "a third component with sharp lines and a constant velocity, close to the poorly determined systemic velocity…" and that if the primary eclipse is total, CF Tau is an evolved system. Our findings agree with his observations. CF Tau was one of the stars studied photometrically by Liakos, Zasche, & Niarchos, P. (2011) , based on unfiltered SWASP photometry, but their study did not include radial velocity measurements as ours does, their photometric results were based on many fewer observations, and their published absolute results differ significantly from ours. We reanalyze their photometric data below. They estimated a minimum mass for the third body, B, as 0.89 ± 0.02 solar masses from the O-C variations. This is rather different from our result of 1.167 ± 0.054 solar masses derived below.

2. ECLIPSE TIMING AND SPECTROSCOPIC STUDY

Spectroscopic observations of CF Tau were obtained at the Harvard-Smithsonian Center for Astrophysics using nearly identical spectrographs on the 1.5-m Tillinghast reflector at the F. L. Whipple Observatory (Mount Hopkins, Arizona), and on the 4.5-m equivalent Multiple Mirror Telescope (also on Mount Hopkins), prior to its conversion to a monolithic 6.5-m telescope. A single echelle order 4.5 nm wide was recorded with intensified photon-counting Reticon detectors, at a central wavelength of about 519 nm containing the lines of the Mg I b triplet. A total of 56 usable observations were obtained from 1985 December to 2009 January at a resolving power of approximately R = 35,000.

The signal-to-noise ratios of these observations range from 22 to 55 per resolution element of 8.5 km/s.

The spectra of CF Tau are clearly triple-lined (see Figure 1), with the sharp lines of the tertiary always lying between the broad lines of the two main components, and generally showing small but noticeable displacements over the years, suggesting physical association. This is confirmed in our detailed analysis below.

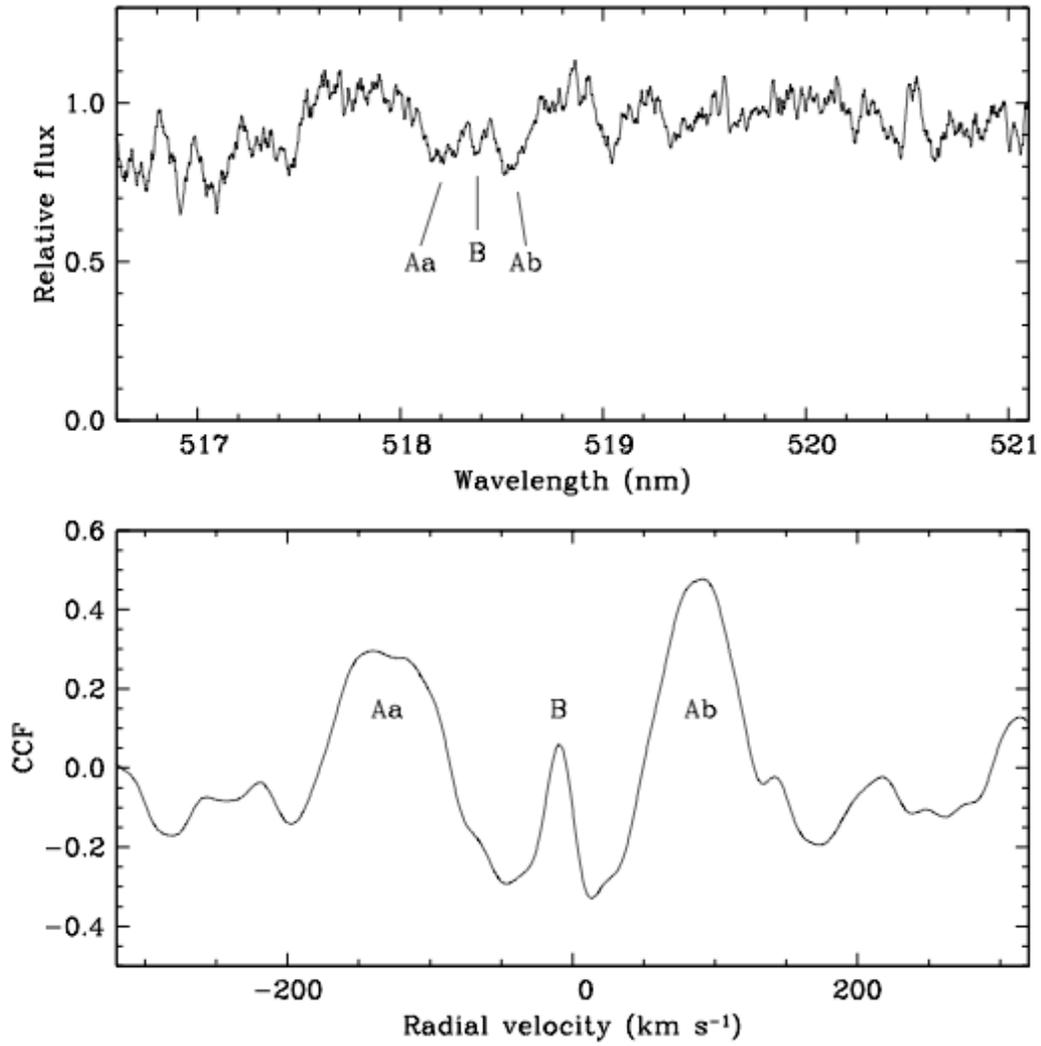

Figure 1. Sample spectrum of CF Tau (HJD 2,446,802.6201) with the corresponding one-dimensional cross-correlation function (CCF) shown in the lower panel. The three

components of the system are indicated: Aa is the cooler and more massive star of the eclipsing pair, and B is the tertiary.

To measure radial velocities for all three components we used an extension of the two-dimensional cross-correlation technique TODCOR (Zucker & Mazeh 1994) to three dimensions (Zucker et al. 1995). Cross-correlations were performed against synthetic templates based on model atmospheres by R. L. Kurucz. The templates for CF Tau were selected from a large library at our disposal in the same way as described recently by Torres et al. (2012), by determining the set of three synthetic spectra (one for each star) giving the best match as measured by the cross-correlation coefficient averaged over all exposures. The template parameters adopted for the velocity determinations correspond to effective temperatures ($T_{eff}$) of 5250 K and 6000 K for the more massive and less massive stars in the eclipsing binary (spectroscopic primary and secondary), 5750 K for the tertiary, and rotational velocities (v sin i when seen in projection) of 60 km/s, 40 km/s, and 2 km/s, respectively. Surface gravities and metallicity (log g and [Fe/H]) have a much smaller impact on the velocities. Solar metallicity was assumed throughout, and for the log g gravities we adopted values of 3.5 and 4.0 for the cool and hot stars of the binary, and 4.0 for the tertiary. The values for the binary are close to those determined from our final analysis in Sect. 4. We improved these stellar parameters by interpolation following the procedures detailed by Torres et al. (2002), and obtained temperatures of 5930 K, 5240 K, and 6060 K, for the spectroscopic primary, secondary, and tertiary (with estimated uncertainties of 150 K), and projected rotational velocities of 56 ± 4 km/s, 38 ± 2 km/s, and 2 ± 2 km/s, respectively. The spectroscopic primary and secondary temperatures are further refined below, incorporating information from the light curve analysis that constrains their difference. We also measured the brightness ratios between the three stars in our spectra (at the mean wavelength of 519 nm) and corrected them to the V band, obtaining L(hot)/L(cool) = 1.16 ± 0.10 and a fractional contribution of the third star to the total light of $L_B$ = 0.16 ± 0.03. These values are consistent with those derived from the light curve analysis in the next section.

The zero point of our velocity system was monitored by taking exposures of the dusk and dawn sky, and small corrections for instrumental shifts were applied to the velocities in each run as described by Latham (1992). Additional adjustments to the velocities were applied to correct for systematic errors resulting from the narrow spectral window and residual blending effects in TODCOR, based on numerical simulations analogous to those described by Torres et al. (2012). The resulting velocities for all three stars are listed in Table 1, where typical uncertainties are 2.8, 2.1, and 1.0 km/s for the spectroscopic primary, secondary, and tertiary.

Inspection of the tertiary velocities as a function of time revealed changes that are correlated with the motion of the center-of-mass of the binary, implying a hierarchical triple system. An initial orbital solution solving for the inner and outer orbits simultaneously gave an estimate of the period of the outer orbit as ~8000 days and a rather high eccentricity, but the phase coverage is incomplete (although a full cycle has elapsed since the beginning of our observations, and periastron passage is covered) and the elements are much more uncertain than those of the inner orbit.

Times of eclipse for CF Tau have been recorded since its discovery, using a variety of techniques. An O-C diagram of the more recent measurements that have better precision shows an obvious light-time effect consistent with the 8000-day period we determined. These timings therefore constrain the outer orbit and complement the velocity observations, so we have used them in a joint solution. Table 2 collects the CCD and photoelectric timing measurements available from the literature that are accurate enough for our purposes. Older timings obtained visually or photographically have typical uncertainties that are one or two orders of magnitude worse, and were thus not considered in our analysis as they do not provide any significant constraint on the O-C amplitude. To the published eclipse times may be added values determined from the ROTSE photometry of the NSVS survey (a fit to the ROTSE photometry was made in the fashion discussed in the next section by using the *jktebop* program, allowing the reference time of eclipse to vary in order to determine precise values of primary and secondary eclipse during the epoch of those measurements). Additional unpublished times of eclipse were determined from the NFO data discussed below by using the methods of Lacy (2006). We incorporated all of these determinations in our orbital fit following the formalism of Irwin (1952), and corrections were applied to the spectroscopic times of observation by iterations to account for the light-time effect on the radial velocities. The joint orbital fit using radial velocities and timings (see Torres 2004) was performed by non-linear least-squares using the Levenberg-Marquardt algorithm (Press et al. 1992), solving for the elements of the inner and outer orbits simultaneously. This obviates the need for a separate fit to the timings in order to model the light-time effect (typically referred to as a "LITE" curve), as this effect follows directly from the elements of the outer orbit, which is constrained in our global fit both by the velocities and the eclipse timings. The orbital period of the inner orbit and the reference epoch of eclipse resulting from our fit constitute the linear ephemeris that we use in our modeling of the differential photometry below. A similar application of this global modeling technique may be seen in the work of Torres et al. (2006). Tests allowing for non-zero eccentricity in the inner orbit gave values indistinguishable from zero, consistent with evidence presented in the next section, so we have assumed for the remainder of the paper that orbit is circular. We also allowed for systematic shifts between the spectroscopic primary (more massive and cooler eclipsing star) and secondary velocity scales, and between the spectroscopic primary and tertiary velocity scales, to account for possible template mismatch in our TODCOR determinations. Both fitted offsets turned out to be small, and do not affect the results. The published uncertainties for eclipse timings are often underestimated, so we rescaled them by iterations in our solution, in order to achieve reduced chi-square values of unity separately for the deeper and shallower eclipse timings. The measurements for eclipses of the hotter star (deeper eclipse) required a scale factor of 2.4, while those of the cooler star (shallower eclipse) were increased by a factor of 4.6, qualitatively consistent with the smaller depth of those eclipses. O-C values in Table 2 are based on the fitted parameters in Table 3. The $\sigma$ and N symbols in Table 3 refer to the standard errors of the residuals to the fits of the spectroscopic orbits and the light-time orbits, and the numbers of observations that were used in fitting these orbits.

The orbital elements from our global fit are listed in Table 3, along with derived quantities. The period of the outer orbit is 8375 ± 136 days (22.93 ± 0.37 yr), which is about half of the period reported by Albayrak et al. (2006). Residuals from this solution are included in Table 1, which also gives the light-time corrections to the spectroscopic dates. Our radial-velocity measurements for the spectroscopic primary and secondary of CF Tau are shown in Figure 2 (corrected for the motion in the outer orbit), along with the computed orbit and the residuals.

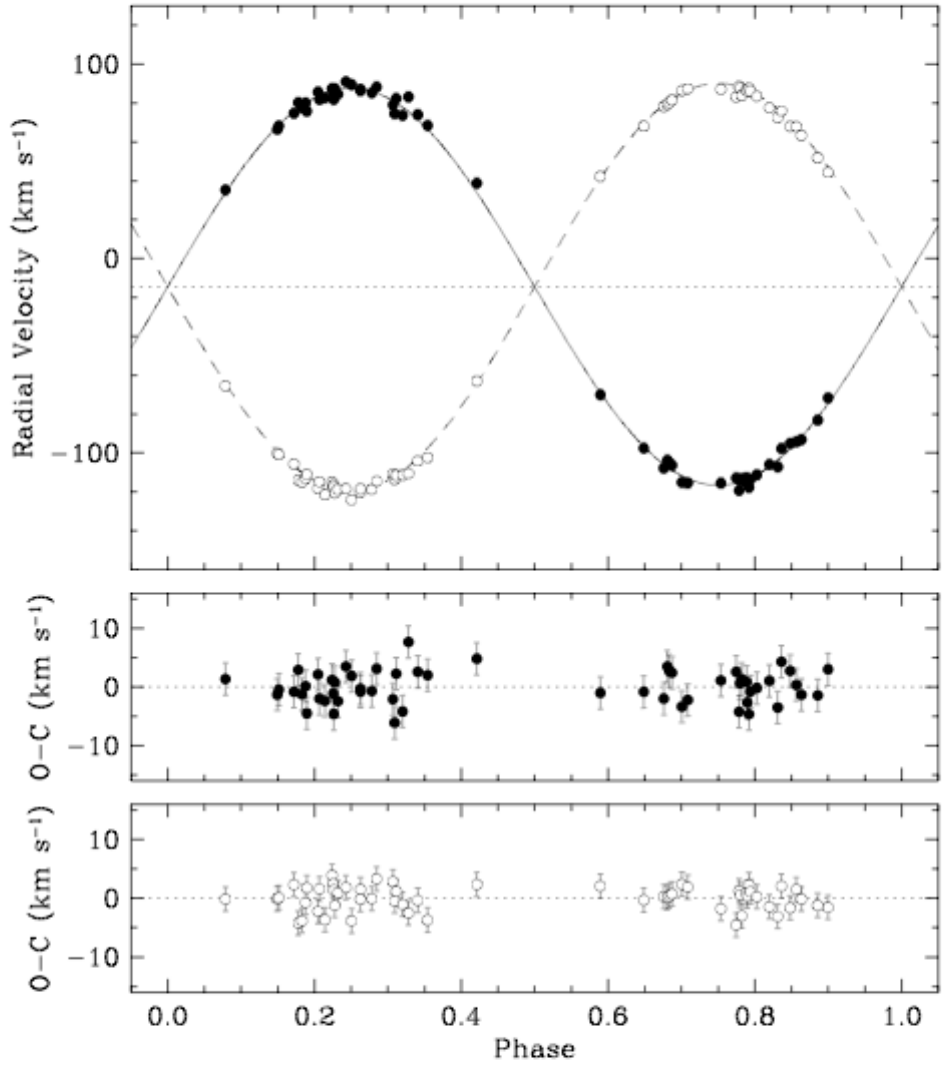

Figure 2. Radial velocity measurements for the spectroscopic primary and secondary of CF Tau along with our best-fit model, after subtraction of the motion in the outer orbit. Phase

0.0 corresponds to the eclipse of the hotter (less massive) star in the binary, represented with open circles. The solid and dashed lines are the computed curves for the cooler and hotter stars, respectively, and the horizontal dotted line corresponds to the center-of-mass velocity of the triple system. Residuals are shown at the bottom.

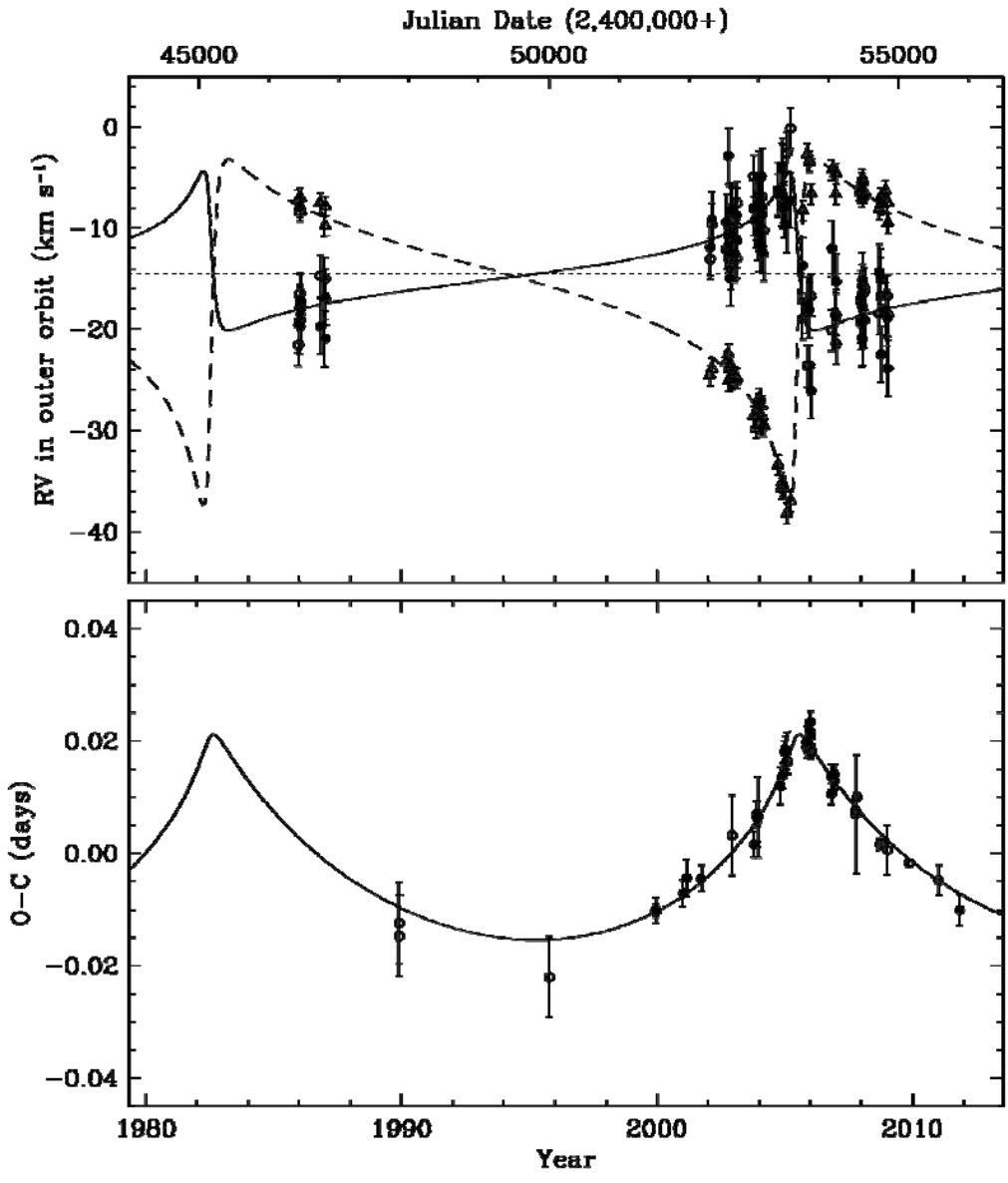

Figure 3. (Top) Radial velocity measurements of CF Tau in the outer orbit. Motion in the inner orbit has been subtracted from the measurements of both components of the eclipsing

pair (shown with open and filled symbols for the hotter and cooler stars, respectively). The velocities of the third star are shown with triangles. The curves correspond to our best-fit model: the solid line represents the motion of the center-of-mass of the eclipsing binary, and the dashed line is the predicted motion for the tertiary. The center-of-mass velocity of the triple system is represented with a horizontal dotted line. (Bottom) O-C diagram, with the primary eclipses (hotter star) represented with open symbols, and secondary eclipses with filled symbols. The curve is the light-time effect predicted by our model.

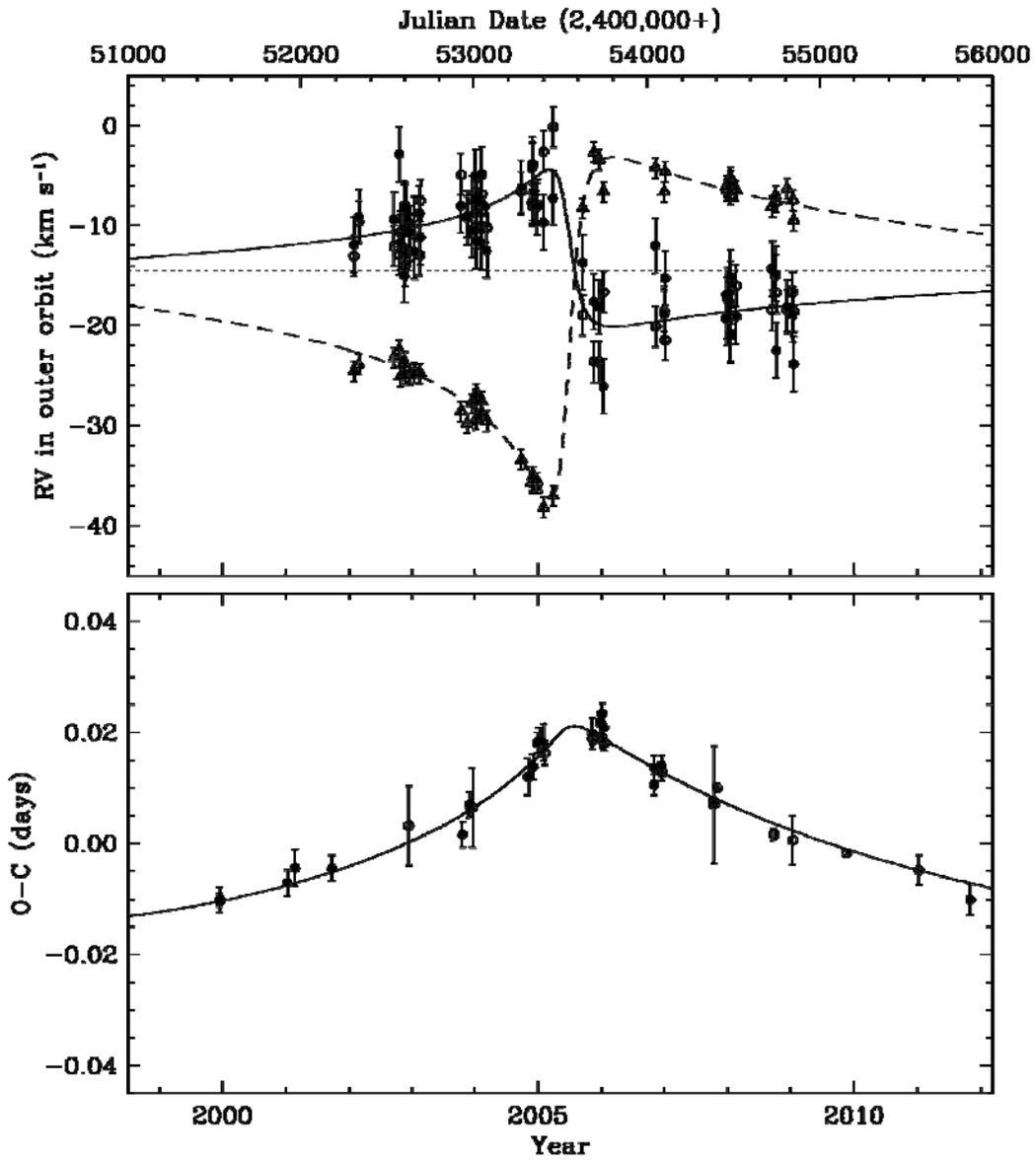

Figure 4. Enlargement of Fig. 3 near the 2005 periastron passage.

Figure 3 (top panel) shows the observations with our model for the outer orbit, where we have subtracted the motion in the inner orbit from the velocities of the primary and secondary. The bottom panel displays the O-C diagram of the eclipse timings from the best-fit linear ephemeris given in Table 3. An enlargement of the 2005 periastron passage is seen in Figure 4.

3. Photometric Study

3.1   Differential photometry

Two independent telescopes were used to measure the brightness of the eclipsing binary star, the URSA WebScope and the NFO WebScope. The URSA WebScope consists of a Meade 10-inch f/6.3 LX-200 telescope with a Santa Barbara Instruments Group ST8 CCD camera (binned 2x2 to produce 765x510 pixel images with 2.3 arcsec square pixels) inside a Technical Innovations Robo-Dome, and controlled automatically by an Apple Macintosh computer. This observatory is located atop Kimpel Hall on the University of Arkansas campus. A Bessell (1990) $V$ filter was used for all observations.

The NFO WebScope is a refurbished 24-inch Group 128 cassegrain reflector with a 2K x 2K Kodak CCD camera, located near Silver City, NM (Grauer, Neely, & Lacy 2008). Observations of 60 s duration through a Bessel (1990) $V$ filter were repeated, often for hours.

Software written by Lacy was used to measure the images (see the description in Lacy et al. 2012). The comparison stars were TYC 1814-0104-1 (V=11.85, G5:, main comparison star, comp) and TYC 1814-0014-1 (V=11.89, G6:, check star, ck). The magnitudes are from Høg et al. 2000, and the spectral types are estimated from those color indices. Both comparison stars are within 6 arcmin of the variable star. The mean nightly comparison star magnitude differences were constant at the level of 0.009 mag (URSA) and 0.017 mag (NFO) for the standard deviation of the mean magnitude differences during a night, and 0.027 mag (URSA) and 0.022 mag (NFO) for the standard deviation of the differential magnitudes during a night. For the differential magnitudes, URSA observations were based only on the comp magnitudes. NFO differential magnitudes were based on the sum of the fluxes of both comparison stars (comp & ck), which was converted to a magnitude called "comparisons". The resulting 8052 (URSA) and 5013 (NFO) $V$ magnitude differences (variable-comp and variable-comparisons) are listed in Tables 4 & 5 (without any nightly corrections) and are shown in Figure 5 (after the nightly corrections discussed below have been added, and a few clearly aberrant points were removed).

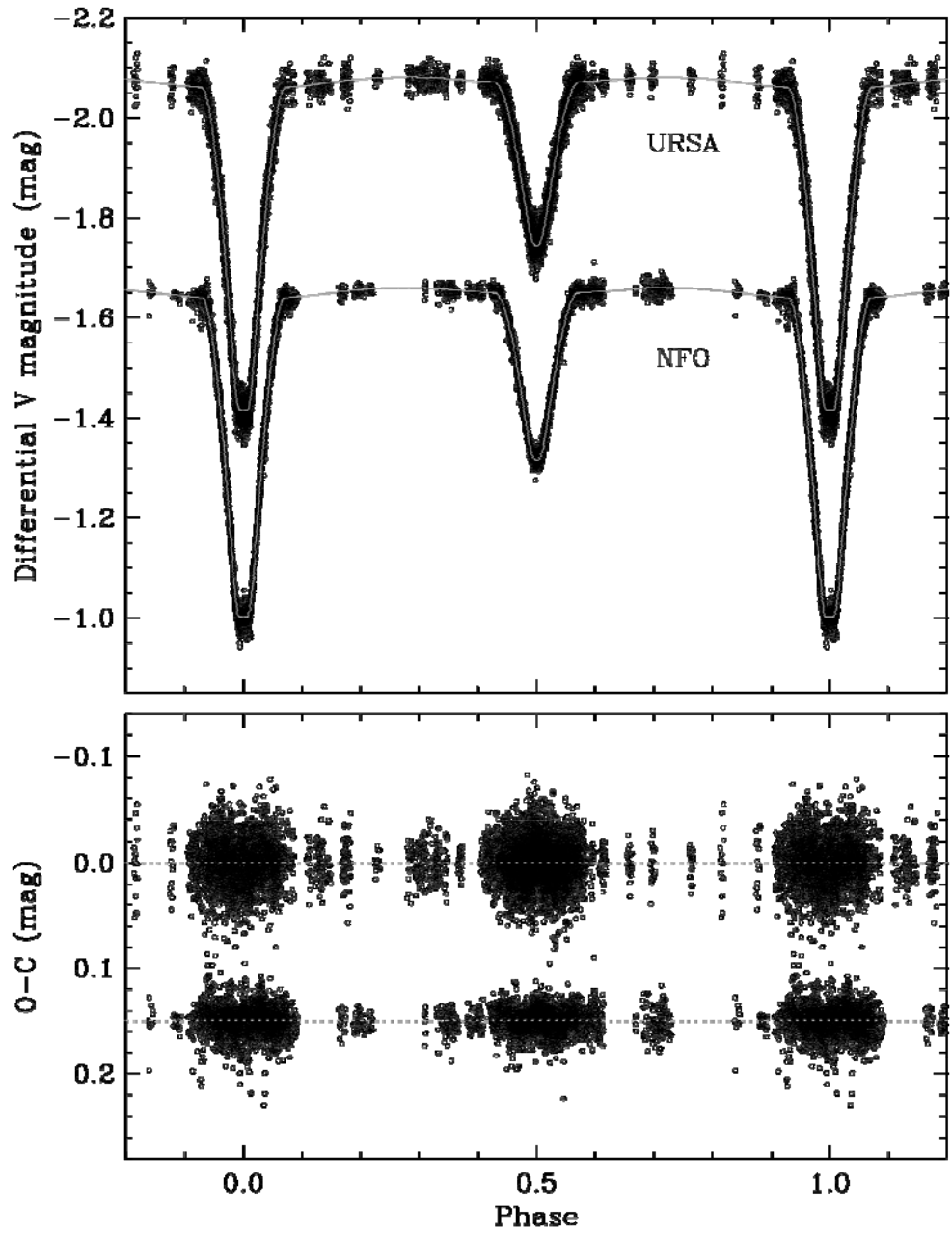

Figure 5. Light Curves of CF Tau from the URSA and NFO WebScope data sets after removal of the nightly corrections. The curves are offset for clarity. Residuals from the fitted photometric orbit are shown at the bottom. Phase zero corresponds to the eclipse of the hotter, less massive star. The orbital phases have been corrected for motion of the eclipsing pair around the center of mass of the triple star system.

Small variations in the zero point of the magnitude system were seen in the NFO measurements, and to a much smaller extent, also in the URSA data. These variations of a few hundredths of a mag are due to sensitivity changes across the field of view of the telescope, coupled with imprecise centering from night to night. The fact that the variations are very much smaller in the URSA data shows that the effect is intrinsic to the telescope design, and not due to night-to-night variations in the stars being measured. Such effects are a well-known feature of wide-field Cassegrain telescopes. They can be partially removed by fitting the variations with a 2D polynomial (see Selman 2004).

From our orbital fit in the preceding section the semi-amplitude of the O-C variations of the eclipse timings corresponds to $27.6 \pm 0.7$ minutes. This is large enough that it has an impact on the light curve fits described below. We have therefore applied appropriate corrections for this light-time effect to the times of observation of our differential photometry.

3.2   Light curve modeling

The light curves were fitted with the Nelson-Davis-Etzel (NDE) model as implemented in the code *jktebop* (Etzel 1981; Popper & Etzel 1981, Southworth et al. 2007), adopting the linear ephemeris:

HJD min I = 2,454,708.90707(66) + 2.75587690(79) E

based on our global fit described earlier (see Table 3). The main adjustable parameters of the eclipsing binary star model (stars Aa and Ab) are the relative central surface brightness of the cooler star Aa ($J_{Aa}$) in units of the central surface brightness of the hotter star Ab, the sum of the relative radii of the cooler and hotter stars ($r_{Aa} + r_{Ab}$) in units of the separation, the ratio of radii ($k = r_{Aa}/r_{Ab}$), the inclination of the orbit ($i$), and the geometric eccentricity factors $e \cos \omega$ and $e \sin \omega$. Auxiliary parameters needed in the analysis include the gravity-brightening exponent, which we adopt as 0.35 for the hotter star Ab and 0.42 for the cooler star Aa based on their temperatures (Claret 1998). For the analysis of the data, quadratic limb-darkening coefficients ($u, u'$) were adopted from the tables of Claret & Bloeman (2011, least-squares method) because a quadratic limb-darkening law produced a better fit to the data than a linear law did. The mass ratio ($q = M_{Aa}/M_{Ab} = 1.024$) was adopted from the spectroscopic analysis in section 2. Two auxialliary parameters (the magnitude at quadrature, between the deeper and shallower eclipses, and the phase of the deeper eclipse) were also fitted. "Third light" $L_B$ was included as a parameter to be fitted by the method. "Reflected light" was calculated from bolometric theory (see Popper & Etzel 1981). The method converged to a solution for both the URSA and NFO data sets (Table 6). In this table, the values of $\sigma$ are the standard deviations of the residuals from the fits, and N are the number of observations used in the fits.

As discussed above, small adjustments were made to remove nightly instrumental variations due to the telescopes' variable responsivity. The number of nights where these adjustments were applied is listed in table 6 as "Corrections." Fits to the "corrected" data then show significantly reduced residual variance, and we have adopted these improved fits (table 6).

The "third light" parameter $L_B$ was found to be significantly non-zero (as expected because of the spectroscopic presence of a third star in the system), and fits to the URSA and NFO data sets did agree well on its value, fitted at 15% of the total light. This is due to the component that is causing the light-time effect as discussed in section 2, the same component that Popper saw in his spectra near the center-of-mass velocity.

Monte Carlo simulations were done where synthetic data were generated and analyzed in the same way as the measured data. In this way we could gauge the accuracy of the uncertainty estimates. 500 simulations were run, allowing us to check the uncertainty's accuracy to about 2 digits. Values of parameter uncertainty generally agreed between the original fitting method and the Monte Carlo method, but when they did not agree exactly, the larger uncertainty estimate was adopted. The uncertainties of the adopted photometric elements in Table 6 take into account both the error estimates from the Monte Carlo method and the degree of agreement between the independent parameter values from the URSA and NFO telescopes. Plots of the fits are shown in Fig. 5.

Lacy (1987) showed that the difference in visual surface brightness parameter, $\Delta F_v$, is related to the normalized $V$-band central surface brightness of the cooler star in eclipsing binaries: $\Delta F_v = 0.25 \log J_c'$. Here, $J_c$ is a parameter, the central surface brightness of the cooler star, that is fitted in the *jktebop* code that we use to model the light curves, and $J_c'$ is the average surface brightness after correction for limb darkening. The relationship between the visual surface brightness parameter $F_v$ and the stellar temperature is tabulated in Popper's (1980) Table 1, thus the temperature difference is readily and very accurately determined from the *V*-filter light curve fit alone. Additional light curves in different bandpasses are not needed in order to determine accurately the temperature difference if this method is used.

The intrinsic color indices and amount of interstellar reddening were determined through a deconvolution procedure using the uvbyβ indices of Lacy (2002). We explored combinations of synthetic Strömgren indices for the three stars taken from the calculations of Girardi et al. (2000), as well as a range of reddening values, seeking to obtain total magnitudes (combined light of the three stars) consistent with the measurements. We imposed the measured light ratios in Table 6 as constraints, along with the condition that the masses of the primary and secondary should be close to those determined in this paper. We obtained a very good fit for a color excess of E(b-y) = 0.21, with a negligible dependence on the metallicity assumed. This corresponds to E(B-V) = 0.28. As a consistency check, we collected all available absolute photometry for CF Tau in the Johnson-Cousins, Strömgren, Tycho-2, and 2MASS systems, and formed 11 non-independent color indices for which color/temperature calibrations exist (Casagrande et al. 2010). We de-reddened each index using E(B-V) = 0.28 and the extinction prescription by Cardelli et al. (1989), and computed the average of the 11 temperatures as 5590 K. This is close to the luminosity-weighted spectroscopic determination of 5650 K (Sect. 2), indicating that the adopted reddening is reasonable.

Tests comparing the NDE model used by *jktebop* with more complicated models (Popper & Etzel 1981; North & Zahn 2004) including the WD (Wilson & Devinney 1971) model, have been made by others. It was found that the limits for high-accuracy determination of parameters such as the radii, inclination, etc. with the NDE model are mean radii less than 0.25 and component oblateness less than 0.04. The CF Tau properties are all within these limits, so we do not feel the need to use a more complicated model in this case.

The photometric parameters of the orbit by Liakos et al. (2011) are compared to our adopted solution in Table 7. The differences in fitted parameter values are likely due to the much smaller numbers of observations in minima in the Liakos et al. solution and the lack of an accurate spectroscopically measured mass ratio by them.

We have obtained and re-analyzed the SWASP data they used. Our results are given in Table 7 (under SWASP) and displayed in Figure 6. The orbital inclination parameter is very poorly determined from the SWASP data set, but when fixed at the mean value of the URSA and NFO results, the other fitted parameters are not significantly different from the URSA and NFO values.

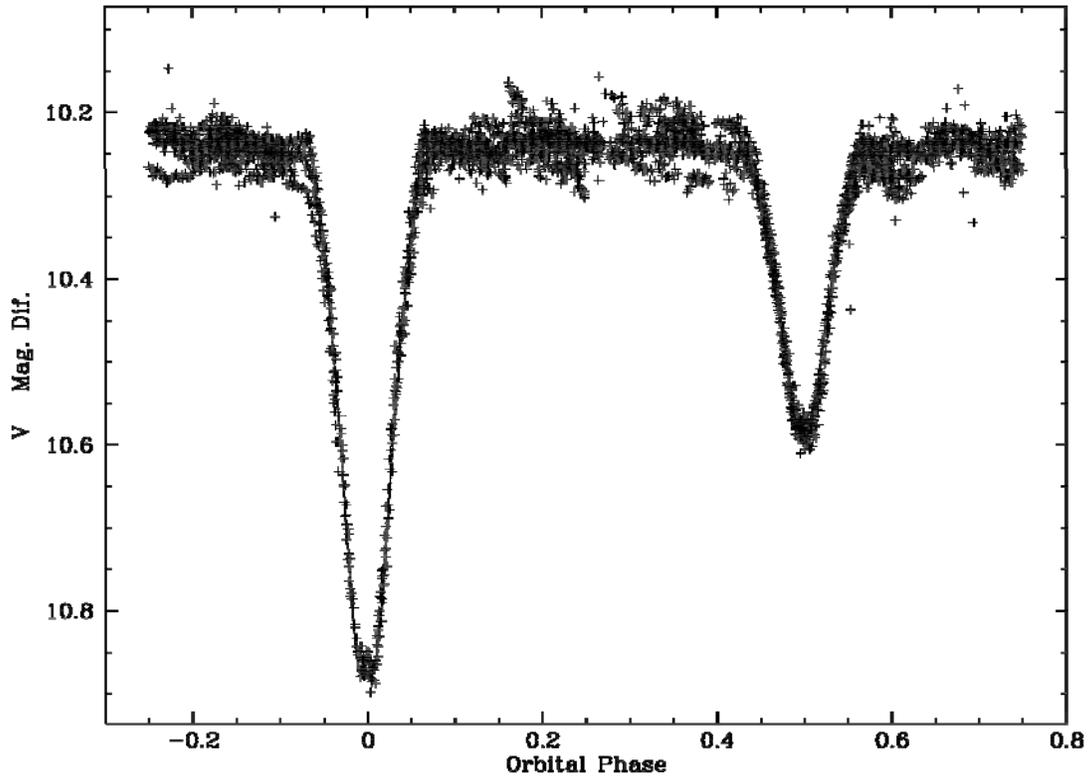

Figure 6. Light cure and orbital fit for the SWASP dataset. The fitted orbital parameters are not significantly different from those of the URSA and NFO datasets, but the number of observations is many fewer and the observational error is greater (see Tables 6 & 7).

4. Absolute Properties

The combination of the spectroscopic results of Table 3 with the light curve results in Table 6 leads to the absolute dimensions and masses for CF Tau shown in Table 8. Table 1 of Popper (1980) has been used for the radiative quantities. The masses are determined to an accuracy of better than 0.9% (standard error), and the radii are good to better than 0.8% (standard error). We have conservatively estimated the uncertainties in the effective temperatures to be 150 K (standard error) to include possible systematic errors in the photometry, spectroscopy, and in the calibrations of Popper (1980). Absolute magnitudes for the stars in the eclipsing pair computed using bolometric corrections are consistent with those directly from the visual surface brightness parameter $F_V$ to within 0.1 mag. The mean distance using bolometric corrections (BCs) is $258 \pm 30$ pc, also consistent with the $F_V$ value. Separate distances for the primary and secondary using BCs are 253 and 262 pc, indicating good internal consistency. The distance we derive for the system corresponds to a parallax of $\pi = 3.73 \pm 0.37$ mas. The Hipparcos catalogue does not include this star.

From the absolute masses in Table 8 and the minimum mass of the binary from our orbital fit in Table 3 we infer an inclination angle for the outer orbit of 57 deg. This leads to a tertiary mass of $1.167 \pm 0.054$ solar masses. Its absolute visual magnitude can be computed based on $L_B$ (from Table 7) and the absolute magnitudes of the other two stars. Adopting for the latter the results based on $F_V$ (Table 8), we obtain a tertiary absolute magnitude of $4.20 \pm 0.14$. Alternately, if we adopt instead the absolute magnitudes of the primary and secondary computed by using bolometric corrections, the result for the tertiary is $4.28 \pm 0.15$. The semimajor axis of the outer orbit is 12.48 AU, which at the distance of the system corresponds to 47 mas. Given the ~2 mag V-band brightness difference between the tertiary and the binary, the wide pair should be fairly easy to resolve with long-baseline interferometry.

5. COMPARISONS WITH THEORY

The accurate dimensions obtained for CF Tau allow for interesting comparisons with stellar evolution theory as well as tidal evolution calculations in close binaries. We discuss these in turn.

5.1 STELLAR EVOLUTION

In Figure 7 we display the measurements for CF Tau in the plane of surface gravity versus effective temperature, against standard evolutionary tracks from the Yale series (Yi et al. 2001; Demarque et al. 2004) interpolated for the exact masses we measure (the mass uncertainty is indicated by the shaded area around each track). These models include convective core overshooting in the amount of $\alpha_{ov} = 0.20$ in units of the pressure scale height. The metallicity has been adjusted to provide the best match to our spectroscopic temperatures. We obtain a satisfactory agreement for [Fe/H] = -0.14, although the temperature difference predicted by theory is seen to be somewhat larger than we measure. The mean age of the system according to these models is 4.3 Gyr.

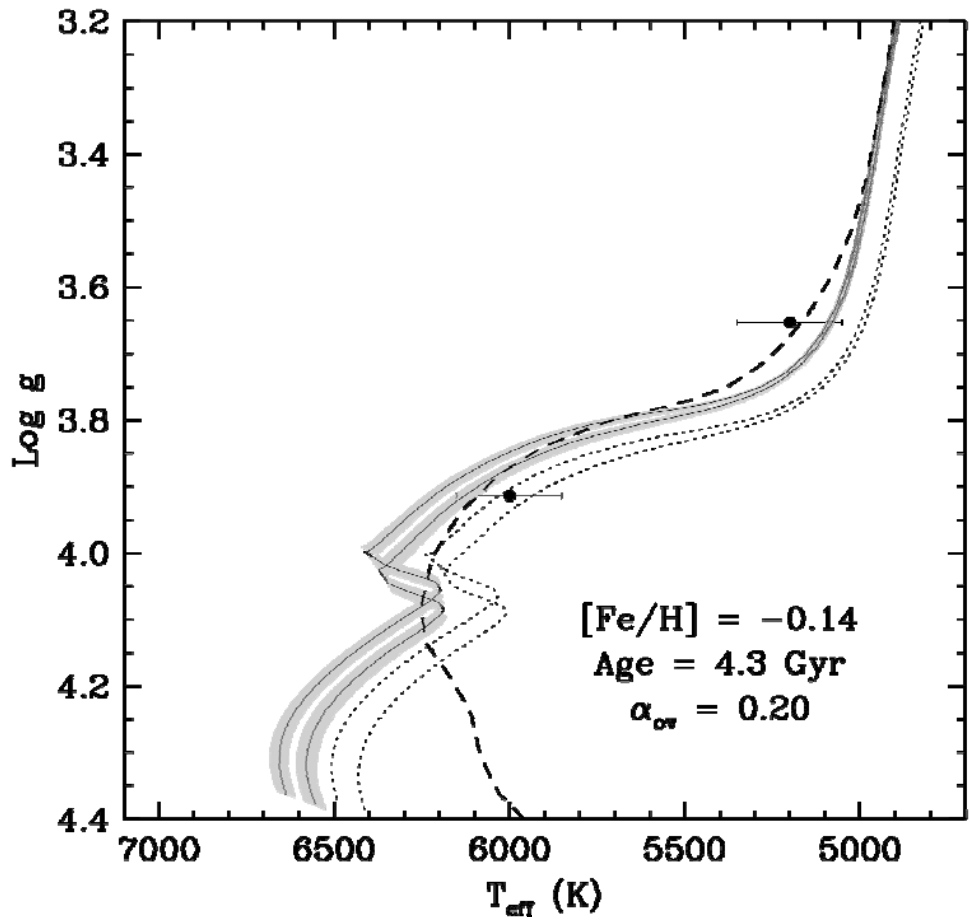

Figure 7. Surface gravity and temperature measurements for the eclipsing components of CF Tau, compared with evolutionary tracks from the Yale series (Yi et al. 2001, Demarque et al.

2004) for the measured masses and $\alpha_{ov}$ = 0.20. The uncertainty in the location of the tracks that comes from the mass errors is indicated by the shaded regions. The metallicity has been adjusted to the value of [Fe/H] = -0.14 that provides the best match. An isochrone for this metallicity and the mean age of the system (4.3 Gyr, according to these models) is represented with a dashed line. Solar-metallicity tracks are shown for reference (dotted lines).

Using the dynamical mass of 1.167 ± 0.054 solar masses that we infer for the tertiary (Sect. 4), an isochrone for the above age and metallicity predicts properties for this star as $M_V$ = 3.69 (+0.36/-0.45), $T_{eff}$ = 6175 (+60/-80) K, and $L_B$ = 0.231 (+0.082/-0.053). All of these quantities are consistent with our measured values, within the uncertainties.

As indicated by the location of the primary and secondary stars in this diagram, CF Tau is a rare case of a well-measured system of nearly equal-mass components (mass difference of only ~2.5%) in very different evolutionary states: one star appears to be crossing the Hertzprung gap, while the other is already at the base of the giant branch. Such systems with significant differential evolution are particularly useful for testing theory, especially when the metallicity has been measured. The best known example is the classic binary AI Phe (Andersen et al. 1988; see also Torres et al. 2010), which is remarkably similar to CF Tau. Unfortunately the chemical composition of CF Tau is unknown. Nevertheless, it may still pose strong constraints on the importance of convective core overshooting.

To explore this, and at the same time to investigate how overshooting affects the temperature difference, we made use of the stellar evolution models of Claret (2004). For comparison purposes we adjusted the metallicity in these models using the same value of $\alpha_{ov}$ = 0.20 as above, and found the best fit for [Fe/H] = -0.12, nearly identical to the previous composition, and a mean age of 4.5 Gyr. We then held the metallicity fixed, and varied the overshooting between $\alpha_{ov}$ = 0.00 and 0.40. The results are shown in Figure 8a-e.

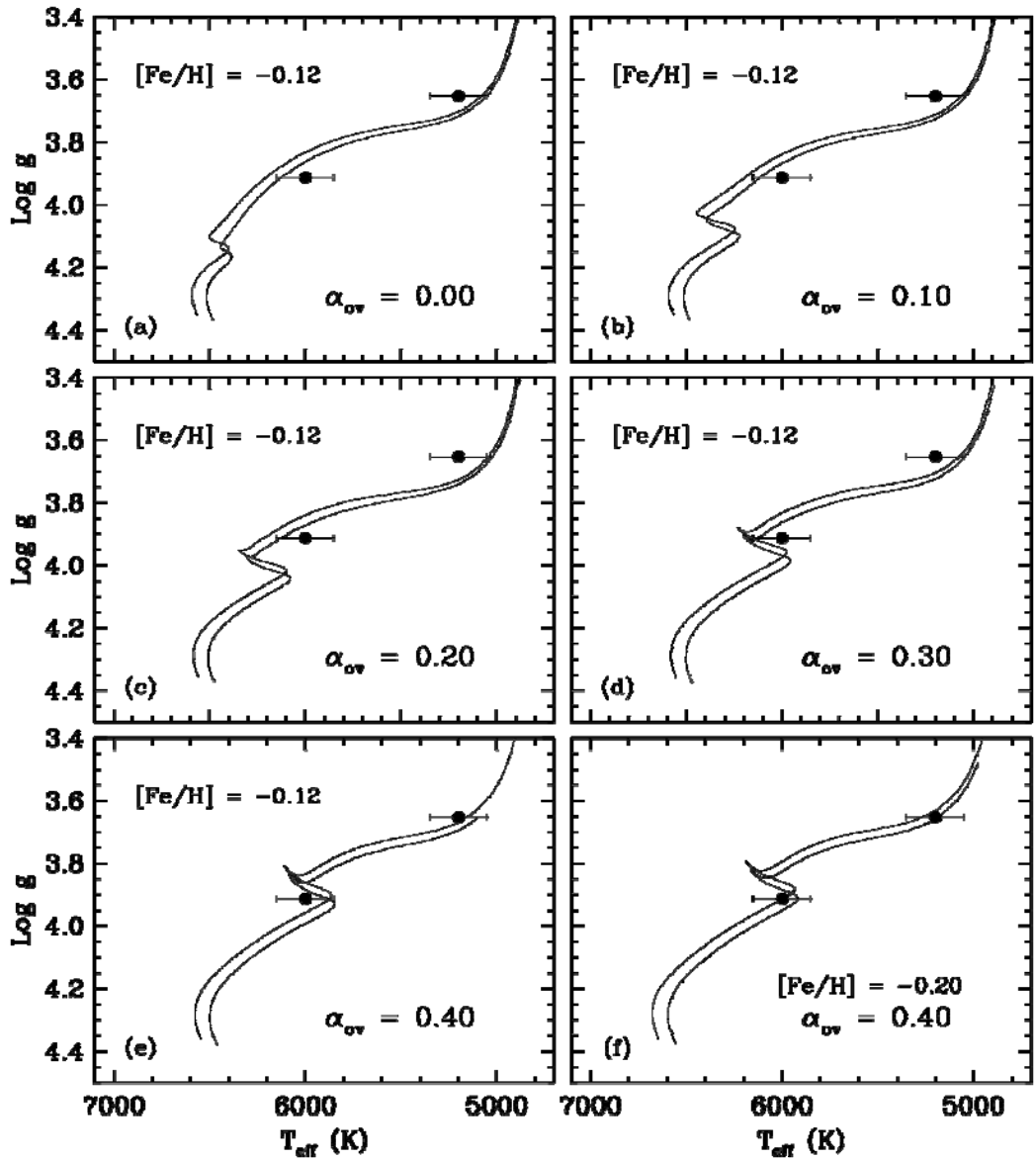

Figure 8. Similar to Fig. 7 showing the eclipsing components, using the models by Claret (2004). (a)-(e) Evolutionary tracks for increasing values of $\alpha_{ov}$ as labeled, and a fixed

metallicity of [Fe/H] = -0.12 that provides the best fit for $\alpha_{ov}$ = 0.20, to allow comparison with a similar fit in Fig. 7. (f) Evolutionary tracks for $\alpha_{ov}$ = 0.40 in which the metallicity has been adjusted to [Fe/H] = -0.20, providing a nearly perfect match to the measured temperatures (see text).

Because overshooting extends the core hydrogen-burning phase, higher values of $\alpha_{ov}$ place the secondary star closer to the end of the main sequence, rather than in the Hertzprung gap. At $\alpha_{ov}$ = 0.40 we find that a small change in the metallicity allows the measured temperatures of the stars to be reproduced almost exactly (Fig. 8f). We note, however, that this high a value of $\alpha_{ov}$ seems at odds with typical values found for other binaries of similar mass (see Claret 2007). An accurate metallicity determination for CF Tau would be highly beneficial, as it would eliminate a free parameter in this modeling.

5.2 TIDAL EVOLUTION

Based on the old age estimated for the CF Tau system, and its short orbital period, the general expectation from theory is that the orbit should be circularized by now due to tidal forces, and that the rotations of the stars should be locked in synchronism with the orbital motion (see, e.g., Mazeh 2008). This is indeed what the observations suggest. To place these predictions on a more quantitative basis we have carried out tidal evolution calculations using the formalism of Hut (1981), integrating the set of six coupled differential equations describing the time evolution of the semimajor axis, eccentricity, angular rotation rates, and inclination of the spin axes relative to the orbital axis. The procedure we follow is the same as that described by Torres et al. (2009), and applied also recently to the BF Dra system by Lacy et al. (2012). Because the initial conditions are unknown a priori, we experimented with different values. The properties of the stars at each time step were taken from the same Claret (2004) evolutionary tracks used in the previous section, for [Fe/H] = -0.12 and $\alpha_{ov}$ = 0.20. Changes in these parameters have a negligible effect on the results.

The outcome of our tidal calculations is displayed graphically in Figure 9. The first two panels show the evolution of the orbital eccentricity and period for two different sets of initial conditions ($P_0$ = 3 days, $e_0$ = 0.3, thin line; and $P_0$ = 4 days, $e_0$ = 0.5, heavy line). The models indicate that regardless of the initial conditions the orbit circularizes just before an age of approximately 1 Gyr, consistent with the observation that the eccentricity at the present age (vertical dotted line) is effectively zero. Figure 9c illustrates the evolution of the angular rotation rates of the two stars (solid lines for the more massive primary, dashed for the secondary), which for convenience we have normalized to the orbital rate. The pseudo-synchronous rotation rate (Hut 1981) is indicated with the dot-dashed lines. We used the same two sets of initial conditions as before, but theory shows once again that the end result does not depend strongly on those values, and pseudo-synchronization is achieved very early on, at an age of about 3 Myr (log age = 6.5). Another effect of tidal forces is that they tend to align the spin axes of the stars with the orbital axis. The evolution of the spin-orbit angles ($\phi$) for CF Tau is illustrated in Figure 9d for the same { $P_0$, $e_0$} values used above. The curves shown were calculated for initial values of $\phi$ = 90 deg for both stars (spin axes perpendicular to the orbital axis), but the results indicate that these angles are quickly damped and reach 0 deg before an age of 10 Myr, much younger than the current age of the system. Smaller initial values of $\phi$ would therefore not change this conclusion.

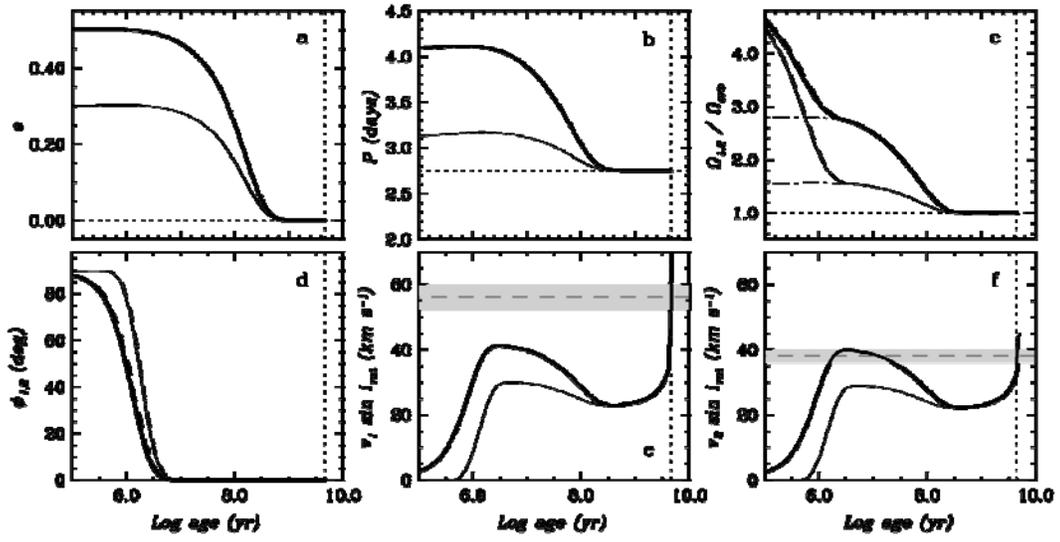

Figure 9. Tidal evolution calculations for the eclipsing components of CF Tau, for two sets of initial conditions regarding the period and eccentricity: $P_0$ = 3 days, $e_0$ = 0.3 (thin line) and $P_0$ = 4 days, $e_0$ = 0.5 (heavy line). The current age of the system (4.5 Gyr, or log age = 9.66 based on the models by Claret (2004), for [Fe/H] = -0.12 and $\alpha_{ov}$ = 0.20) is indicated with a vertical dotted line. (a)-(b) Time evolution of the eccentricity and period. The observed values are marked with horizontal dotted lines. (c) Angular rotation rates, normalized to the orbital rate. Initial values are arbitrarily set to $\Omega/\Omega_{orb}$ = 5.0. The curves for the more massive primary (solid) and secondary (dashed) are nearly indistinguishable. Pseudo-synchronous rates as a function of time are indicated with the dot-dashed lines. (d) Spin-orbit inclination angles, set initially to 90 deg. (e)-(f) Evolution of the projected rotational velocities. Measured v sin i values and uncertainties are indicated with horizontal dashed lines and shaded areas.

Finally, the evolution of the projected rotational velocities of the stars is illustrated in Figure 9e and 9f, with the same initial conditions as above. The measured values for CF Tau are indicated with the dashed lines. While v sin i evolves considerably and is sensitive to initial conditions in the early stages, theory predicts final values at the current mean age of 4.5 Gyr of 51.3 km/s for the more massive star and 37.6 km/s for the other. The latter agrees nearly perfectly with our measurement, and the predicted value for the cooler star is only marginally lower than observed, given our uncertainties.

In conclusion, theoretical predictions from tidal evolution are in good agreement with the observations regarding circularization of the orbit and synchronization of the stellar rotations, and the models indicate also that the spin axes of both stars should be aligned with the orbit (although this is not directly measurable for CF Tau).

ACKNOWLEDGMENTS

The authors wish to thank Bill Neely who operates and maintains the NFO WebScope for the Consortium, and who handles preliminary processing of the images and their distribution. We are grateful to P. Berlind, M. Calkins, R. Davis, G. Esquerdo, T. Fleming, E. Horine, D. Latham, L. Marshall, R. Mathieu, J. Peters, and R. Stefanik for help in obtaining the spectroscopic observations used here. GT acknowledges partial support for this work from NSF grant AST-10-07992.

FIGURE CAPTIONS

Figure 1: Sample spectrum of CF Tau (HJD 2,446,802.6201) with the corresponding one-dimensional cross-correlation function (CCF) shown in the lower panel. The three components of the system are indicated: Aa is the cooler and more massive star of the eclipsing pair, and B is the tertiary.

Figure 2: Radial velocity measurements for the spectroscopic primary and secondary of CF Tau along with our best-fit model, after subtraction of the motion in the outer orbit. Phase 0.0 corresponds to the eclipse of the hotter (less massive) star in the binary, represented with open circles. The solid and dashed lines are the computed curves for the cooler and hotter stars, respectively, and the horizontal dotted line corresponds to the center-of-mass velocity of the triple system. Residuals are shown at the bottom.

Figure 3: (Top) Radial velocity measurements of CF Tau in the outer orbit. Motion in the inner orbit has been subtracted from the measurements of both components of the eclipsing pair (filled and open circles for the cooler and hotter star, respectively). The velocities of the third star are shown with triangles, and the curves correspond to our best-fit model. The center of mass of the triple system is represented with a dotted line. (Bottom) O-C timing residuals (filled symbols for the cooler star) along with our best-fit model of the light-time effect.

Figure 4: Enlargement of Fig. 3 near the 2005 periastron passage.

Figure 5: Light Curves of CF Tau from the URSA and NFO WebScope data sets after removal of the nightly corrections. The curves are offset for clarity. Residuals from the fitted photometric orbit are shown at the bottom. The orbital phases have been corrected for motion of the eclipsing pair around the center of mass of the triple star system.

Figure 6: Light cure and orbital fit for the SWASP dataset. The fitted orbital parameters are not significantly different from those of the URSA and NFO datasets, but the number of observations is many fewer and the observational error is greater (see Tables 6 & 7).

Figure 7: Surface gravity and temperature measurements for the eclipsing components of CF Tau, compared with evolutionary tracks from the Yale series (Yi et al. 2001, Demarque et al. 2004) for the measured masses and $\alpha_{ov}$ = 0.20. The uncertainty in the location of the tracks that comes from the mass errors is indicated by the shaded regions. The metallicity has been adjusted to the value

of [Fe/H] = -0.14 that provides the best match. An isochrone for this metallicity and the mean age of the system (4.3 Gyr, according to these models) is represented with a dashed line. Solar-metallicity tracks are shown for reference (dotted lines).

Figure 8: Similar to Fig. 6 showing the eclipsing components, using the models by Claret (2004). (a)-(e) Evolutionary tracks for increasing values of $\alpha_{ov}$ as labeled, and a fixed metallicity of [Fe/H] = -0.12 that provides the best fit for $\alpha_{ov}$ = 0.20, to allow comparison with a similar fit in Fig. 4. (f) Evolutionary tracks for $\alpha_{ov}$ = 0.40 in which the metallicity has been adjusted to [Fe/H] = -0.20, providing a nearly perfect match to the measured temperatures (see text).

Figure 9: Tidal evolution calculations for the eclipsing components of CF Tau, for two sets of initial conditions regarding the period and eccentricity: $P_0$ = 3 days, $e_0$ = 0.3 (thin line) and $P_0$ = 4 days, $e_0$ = 0.5 (heavy line). The current age of the system (4.5 Gyr, or log age = 9.66 based on the models by Claret (2004), for [Fe/H] = -0.12 and $\alpha_{ov}$ = 0.20) is indicated with a vertical dotted line. (a)-(b) Time evolution of the eccentricity and period. The observed values are marked with horizontal dotted lines. (c) Angular rotation rates, normalized to the orbital rate. Initial values are arbitrarily set to $\Omega/\Omega_{orb}$ = 5.0. The curves for the primary (solid) and secondary (dashed) are nearly indistinguishable. Pseudo-synchronous rates as a function of time are indicated with the dot-dashed

lines. (d) Spin-orbit inclination angles, set initially to 90 deg. (e)-(f) Evolution of the projected rotational velocities. Measured v sin i values and uncertainties are indicated with horizontal dashed lines and shaded areas.

Table 1. Radial velocity measurements for CF Tau

| Year | HJD-2,400,000 | Δt (days) | Inner phase | Outer phase | RV cool (km/s) | RV hot (km/s) | Rv ter (km/s) | O-C cool (km/s) | O-C hot (km/s) | O-C ter (km/s) |
|---|---|---|---|---|---|---|---|---|---|---|
| 1985.98 | 46423.6427 | -0.0028 | 0.6831 | 0.1506 | 73.99 | -118.53 | -7.20 | -1.24 | -3.79 | 0.52 |
| 1986.05 | 46448.5684 | -0.0026 | 0.7278 | 0.1536 | 83.95 | -123.96 | -7.94 | 0.79 | -1.18 | -0.13 |
| 1986.05 | 46449.5655 | -0.0025 | 0.0896 | 0.1537 | -73.52 | 38.80 | -7.82 | -1.00 | 2.09 | -0.01 |
| 1986.05 | 46449.8035 | -0.0025 | 0.1760 | 0.1537 | -111.27 | 74.68 | -8.32 | -1.98 | 0.29 | -0.51 |
| 1986.06 | 46453.6693 | -0.0025 | 0.5787 | 0.1542 | 31.85 | -68.96 | -7.03 | 1.34 | -0.13 | 0.80 |
| 1986.82 | 46729.8827 | 0.0005 | 0.8069 | 0.1872 | 76.03 | -113.92 | -7.53 | -2.08 | 2.85 | 1.18 |
| 1987.02 | 46802.6201 | 0.0013 | 0.2007 | 0.1959 | -118.12 | 83.43 | -9.76 | -3.35 | 2.40 | -0.84 |
| 1987.02 | 46805.5941 | 0.0013 | 0.2799 | 0.1962 | -117.21 | 84.76 | -7.86 | 0.63 | 0.58 | 1.07 |
| 2002.09 | 52306.6139 | 0.0038 | 0.3858 | 0.8530 | -79.75 | 55.27 | -24.55 | -1.42 | -1.21 | -1.94 |
| 2002.16 | 52332.5900 | 0.0035 | 0.8114 | 0.8561 | 85.80 | -108.04 | -23.90 | 2.27 | 1.18 | -1.15 |
| 2002.72 | 52537.9287 | 0.0010 | 0.3199 | 0.8807 | -101.94 | 81.60 | -23.24 | 1.06 | -1.42 | 0.73 |
| 2002.80 | 52566.8885 | 0.0006 | 0.8281 | 0.8841 | 87.28 | -106.40 | -22.48 | 7.65 | -2.50 | 1.69 |
| 2002.82 | 52573.8754 | 0.0005 | 0.3634 | 0.8850 | -89.07 | 67.49 | -25.07 | -1.32 | -0.12 | -0.85 |
| 2002.88 | 52595.7261 | 0.0002 | 0.2920 | 0.8876 | -113.59 | 91.86 | -23.59 | -4.62 | 2.36 | 0.78 |
| 2002.88 | 52596.8639 | 0.0002 | 0.7049 | 0.8877 | 89.85 | -114.20 | -23.79 | 2.14 | -2.21 | 0.59 |
| 2002.90 | 52602.7871 | 0.0001 | 0.8542 | 0.8884 | 72.71 | -98.21 | -24.92 | 2.01 | -3.69 | -0.50 |
| 2002.97 | 52629.7878 | -0.0003 | 0.6515 | 0.8916 | 72.51 | -96.54 | -24.93 | -0.45 | 0.11 | -0.31 |
| 2003.05 | 52657.5686 | -0.0007 | 0.7319 | 0.8949 | 88.91 | -114.32 | -24.67 | -2.43 | 0.96 | 0.16 |
| 2003.13 | 52686.6442 | -0.0011 | 0.2822 | 0.8984 | -108.77 | 88.35 | -24.86 | 1.35 | -2.96 | 0.19 |
| 2003.14 | 52690.6221 | -0.0012 | 0.7256 | 0.8989 | 89.78 | -112.10 | -24.81 | -1.13 | 2.51 | 0.27 |
| 2003.78 | 52924.8712 | -0.0050 | 0.7240 | 0.9269 | 92.97 | -109.50 | -28.57 | 1.18 | 3.91 | -1.24 |
| 2003.88 | 52960.8476 | -0.0057 | 0.7782 | 0.9312 | 91.07 | -113.00 | -29.80 | -0.68 | -0.03 | -2.04 |

| | | | | | | | | | |
|---|---|---|---|---|---|---|---|---|---|
| 2003.95 | 52986.8371 | -0.0062 | 0.2086 | 0.9343 | -109.57 | 93.27 | -27.81 | -2.21 | 1.94 | 0.28 |
| 2004.00 | 53007.6010 | -0.0066 | 0.7429 | 0.9367 | 97.02 | -112.39 | -27.39 | 3.50 | 1.83 | 0.98 |
| 2004.02 | 53011.7650 | -0.0067 | 0.2538 | 0.9372 | -109.52 | 93.20 | -29.38 | 1.10 | -1.77 | -0.95 |
| 2004.02 | 53014.6200 | -0.0068 | 0.2897 | 0.9376 | -110.14 | 92.15 | -26.81 | -2.68 | 0.37 | 1.66 |
| 2004.10 | 53043.6409 | -0.0074 | 0.8201 | 0.9410 | 79.95 | -105.14 | -27.61 | -4.17 | -1.03 | 1.27 |
| 2004.11 | 53044.6422 | -0.0074 | 0.1834 | 0.9412 | -98.88 | 86.59 | -28.70 | 2.73 | 0.41 | 0.19 |
| 2004.11 | 53047.6885 | -0.0074 | 0.2887 | 0.9415 | -106.52 | 93.11 | -27.58 | 0.88 | 0.96 | 1.36 |
| 2004.18 | 53072.6090 | -0.0080 | 0.3312 | 0.9445 | -100.72 | 79.00 | -29.55 | -3.50 | -3.07 | -0.23 |
| 2004.74 | 53274.9816 | -0.0129 | 0.7626 | 0.9687 | 95.35 | -111.82 | -33.41 | -0.29 | -0.16 | 0.03 |
| 2004.90 | 53333.9210 | -0.0146 | 0.1487 | 0.9757 | -88.42 | 77.32 | -35.68 | -0.82 | -0.27 | -0.60 |
| 2004.91 | 53339.8572 | -0.0148 | 0.3027 | 0.9764 | -102.13 | 92.74 | -35.12 | -0.17 | 0.28 | 0.13 |
| 2004.92 | 53340.8912 | -0.0148 | 0.6779 | 0.9765 | 89.42 | -104.77 | -35.49 | 2.92 | -4.19 | -0.21 |
| 2004.99 | 53365.7953 | -0.0156 | 0.7143 | 0.9795 | 92.21 | -111.82 | -35.68 | -2.37 | -3.63 | 0.32 |
| 2005.08 | 53401.6559 | -0.0167 | 0.7263 | 0.9838 | 91.84 | -107.75 | -38.20 | -4.60 | 1.47 | -1.27 |
| 2005.23 | 53456.6258 | -0.0186 | 0.6721 | 0.9904 | 84.85 | -95.63 | -37.03 | -0.81 | 2.35 | 0.11 |
| 2005.71 | 53628.9115 | -0.0209 | 0.1870 | 0.0109 | -109.71 | 78.25 | -8.26 | 2.44 | 0.75 | -0.26 |
| 2005.88 | 53693.8490 | -0.0199 | 0.7506 | 0.0187 | 84.41 | -129.27 | -2.67 | 1.89 | -3.84 | 1.60 |
| 2005.96 | 53722.8514 | -0.0194 | 0.2746 | 0.0221 | -118.26 | 77.80 | -3.43 | 2.58 | -4.54 | 0.22 |
| 2006.03 | 53748.7975 | -0.0189 | 0.6896 | 0.0252 | 70.35 | -116.60 | -6.66 | -4.49 | 1.81 | -3.30 |
| 2006.86 | 54049.8209 | -0.0135 | 0.9212 | 0.0612 | 33.72 | -68.09 | -4.25 | 4.82 | 2.37 | 0.05 |
| 2007.00 | 54100.7446 | -0.0126 | 0.3998 | 0.0673 | -76.62 | 39.48 | -6.67 | 3.02 | -1.53 | -2.07 |
| 2007.01 | 54104.7158 | -0.0125 | 0.8408 | 0.0677 | 69.09 | -109.02 | -4.61 | 2.58 | -0.32 | 0.01 |
| 2007.98 | 54458.8611 | -0.0072 | 0.3482 | 0.1100 | -99.23 | 63.94 | -6.08 | 2.73 | -1.61 | 0.31 |
| 2007.99 | 54462.7595 | -0.0071 | 0.7628 | 0.1105 | 82.38 | -122.57 | -6.46 | -0.80 | 1.53 | -0.05 |
| 2008.04 | 54481.7377 | -0.0069 | 0.6494 | 0.1128 | 62.50 | -104.24 | -5.18 | -1.30 | -0.07 | 1.31 |
| 2008.05 | 54483.6308 | -0.0068 | 0.3363 | 0.1130 | -101.80 | 71.98 | -6.95 | 4.30 | 2.09 | -0.45 |
| 2008.05 | 54484.6518 | -0.0068 | 0.7068 | 0.1131 | 77.84 | -118.93 | -5.53 | -1.97 | 1.63 | 0.98 |
| 2008.14 | 54516.7569 | -0.0064 | 0.3566 | 0.1169 | -98.25 | 63.89 | -6.53 | 0.35 | 1.54 | 0.11 |
| 2008.70 | 54722.9605 | -0.0038 | 0.1808 | 0.1416 | -107.33 | 75.71 | -8.12 | 3.48 | 0.11 | -0.67 |

| 2008.76 | 54743.9156 | -0.0035 | 0.7847 | 0.1441 | 84.72 | -118.13 | -7.01 | 3.11 | 3.32 | 0.52 |
| 2008.77 | 54748.0323 | -0.0034 | 0.2785 | 0.1446 | -122.87 | 85.00 | -7.62 | -4.23 | 1.29 | -0.08 |
| 2008.94 | 54809.7892 | -0.0027 | 0.6879 | 0.1519 | 76.59 | -116.67 | -6.32 | 0.13 | -0.71 | 1.44 |
| 2009.03 | 54841.7735 | -0.0023 | 0.2939 | 0.1557 | -117.04 | 82.71 | -7.45 | -0.79 | 1.14 | 0.42 |
| 2009.05 | 54848.7056 | -0.0023 | 0.8093 | 0.1566 | 71.10 | -117.01 | -9.51 | -6.08 | -0.44 | -1.62 |

Note: ∆t is the correction applied to the dates of observation for light-travel time in the outer orbit.

Table 2. Times of eclipse for CF Tau

| Year | HJD-2,400,000 | Err (days) | Cycle | Typ | Method | O-C (days) | Ref |
|---|---|---|---|---|---|---|---|
| 1989.92 | 47864.4441 | 0.003 | -1395.0 | 1 | pe | -0.0051 | 1 |
| 1989.92 | 47864.4464 | 0.003 | -1395.0 | 1 | pe | -0.0028 | 1 |
| 1995.79 | 50008.5090 | 0.0030 | -617.0 | 1 | ccd | -0.0067 | 2 |
| 1999.96 | 51529.7649 | 0.0005 | -65.0 | 1 | ROTSE | 0.0002 | 3 |
| 1999.96 | 51531.1428 | 0.0005 | -64.5 | 2 | ROTSE | 0.0002 | 3 |
| 2001.03 | 51919.7246 | 0.0005 | 76.5 | 2 | ccd | 0.0005 | 4 |
| 2001.15 | 51966.5772 | 0.0007 | 93.5 | 2 | ccd | 0.0028 | 4 |
| 2001.74 | 52178.7796 | 0.0005 | 170.5 | 2 | ccd | 0.0007 | 5 |
| 2002.95 | 52621.1055 | 0.003 | 331.0 | 1 | ccd | 0.0030 | 6 |
| 2003.80 | 52933.8960 | 0.0005 | 444.5 | 2 | ccd | -0.0036 | 7 |
| 2003.91 | 52972.4836 | 0.0005 | 458.5 | 2 | pe | 0.0010 | 8 |
| 2003.93 | 52979.3728 | 0.0005 | 461.0 | 1 | pe | 0.0004 | 8 |
| 2003.96 | 52993.1522 | 0.003 | 466.0 | 1 | ccd | 0.0002 | 9 |
| 2004.84 | 53311.4615 | 0.0007 | 581.5 | 2 | pe | -0.0019 | 8 |
| 2004.92 | 53341.7779 | 0.0005 | 592.5 | 2 | ccd | -0.0011 | 10 |
| 2004.99 | 53366.5851 | 0.0004 | 601.5 | 2 | ccd | 0.0025 | 10 |
| 2005.02 | 53377.6090 | 0.0005 | 605.5 | 2 | ccd | 0.0025 | 10 |
| 2005.08 | 53399.6557 | 0.0007 | 613.5 | 2 | ccd | 0.0015 | 10 |
| 2005.10 | 53409.2994 | 0.0009 | 617.0 | 1 | ccd | -0.0007 | 11 |
| 2005.85 | 53683.5127 | 0.0006 | 716.5 | 2 | ccd | -0.0002 | 12 |

| Year | JD | Err | Cycle | Typ | Method | O-C | Ref |
|---|---|---|---|---|---|---|---|
| 2005.86 | 53684.8897 | 0.0004 | 717.0 | 1 | ccd | -0.0012 | 10 |
| 2005.98 | 53727.6087 | 0.0004 | 732.5 | 2 | ccd | 0.0025 | 10 |
| 2006.01 | 53738.6338 | 0.0004 | 736.5 | 2 | ccd | 0.0043 | 13 |
| 2006.02 | 53742.7627 | 0.0005 | 738.0 | 1 | ccd | -0.0005 | 13 |
| 2006.04 | 53749.6548 | 0.0005 | 740.5 | 2 | ccd | 0.0020 | 13 |
| 2006.05 | 53753.7858 | 0.0006 | 742.0 | 1 | ccd | -0.0008 | 13 |
| 2006.84 | 54041.7675 | 0.0004 | 846.5 | 2 | ccd | -0.0030 | 13 |
| 2006.84 | 54041.7705 | 0.0005 | 846.5 | 2 | ccd | 0.0000 | 13 |
| 2006.95 | 54084.4860 | 0.0007 | 862.0 | 1 | ccd | 0.0002 | 14 |
| 2006.96 | 54085.8649 | 0.0004 | 862.5 | 2 | ccd | 0.0012 | 13 |
| 2007.78 | 54384.8710 | 0.0004 | 971.0 | 1 | ccd | -0.0008 | 15 |
| 2007.78 | 54387.6264 | 0.0044 | 972.0 | 1 | ccd | -0.0012 | 16 |
| 2007.84 | 54406.9206 | 0.0003 | 979.0 | 1 | ccd | 0.0021 | 15 |
| 2008.73 | 54734.8615 | 0.0004 | 1098.0 | 1 | ccd | -0.0020 | 15 |
| 2009.03 | 54842.3397 | 0.0019 | 1137.0 | 1 | ccd | -0.0017 | 17 |
| 2009.88 | 55153.7515 | 0.0003 | 1250.0 | 1 | ccd | -0.0007 | 18 |
| 2011.03 | 55572.6417 | 0.0011 | 1402.0 | 1 | ccd | 0.0001 | 19 |
| 2011.84 | 55868.8932 | 0.0006 | 1509.5 | 2 | ccd | -0.0029 | 20 |

Notes: The cycle numbers refer to the reference time of eclipse of the hotter star given in Table 3. Internal errors under the "Err" column are reported here as published, but were re-scaled in our orbital solution as described in the text. The type of eclipse under "Typ" is 1 for the deeper eclipse of the hotter star and 2 for an eclipse of the cooler star. The observational technique is indicated under "Method" as 'pe' (photoelectric), 'ccd', or ROTSE. Residuals from our combined orbital fit are listed under "O-C".

References: (1) Hubscher (1990); (2) Paschke (1996); (3) This paper (ROTSE); (4) Lacy et al. (2001); (5) Lacy et al. (2002); (6) Nagai (2003); (7) Lacy (2003); (8) Albayrak et al. (2005); (9) Nagai (2004); (10) Lacy (2006); (11) Hubscher et al. (2005); (12)

Hubscher et al. (2006); (13) Lacy (2007); (14) Hubscher & Walter (2007); (15) Lacy (2009); (16) Hubscher et al. (2008); (17) Hubscher et al. (2010); (18) This paper (NFO); (19) Diethelm (2011); (20) Diethelm (2012).

Table 3. Spectroscopic orbital solution incorporating eclipse timings

| Inner Orbit | |
|---|---|
| P(A) (days) | 2.75587690 ± 0.00000079 |
| γ (km/s) | -14.51 ± 0.39 |
| K(Aa) (km/s) | 102.15 ± 0.42 |
| K(Ab) (km/s) | 104.65 ± 0.32 |
| e | 0 (fixed) |
| Tmin (Ab) (HJD) | 2,454,708.90707 ± 0.00066 (eclipse of the hotter star) |
| offset2 (km/s) | +1.15 ± 0.48 |
| | |

| Outer Orbit | |
| --- | --- |
| | |
| P(AB) (days) | 8375 ± 136 |
| K(A) (km/s) | 7.88 ± 0.25 |
| K(B) (km/s) | 17.10 ± 0.38 |
| e(AB) | 0.8388 ± 0.0065 |
| ω(A) (degrees) | 70.0 ± 1.6 |
| offset3 (km/s) | +0.87 ± 0.48 |
| | |
| Derived Quantities | |
| | |
| K(O-C) (days) | 0.01908 ± 0.00054 |
| M(Aa) $\sin^3 i$ (solar masses) | 1.2780 ± 0.0094 |
| M(Ab) $\sin^3 i$ (solar masses) | 1.247 ± 0.011 |
| q=M(Ab)/M(Aa) | 0.9761 ± 0.0050 |
| a sin i (solar radii) | 11.265 ± 0.029 |
| M(A) sin$^3$ i (solar masses) | 1.496 ± 0.056 |
| M(B) sin$^3$ i (solar masses) | 0.689 ± 0.032 |
| | |
| σ RV(Aa) (km/s) | 2.74 |
| σ RV(Ab) (km/s) | 2.06 |
| σ RV(B) (km/s) | 1.01 |
| N RVs (Aa,Ab,B) | 56 |
| N (Min Aa) | 19 |
| N (Min Ab) | 19 |

Note: offset2 is a systematic offset between the primary and secondary velocities (see text), and offset3 is the same between the primary and tertiary.

Table 4. URSA Differential Photometry of CF Tau

| HJD-2,400,000 | ΔV |
|---|---|
| 51919.66146 | -1.843 |
| 51919.66248 | -1.853 |
| 51919.66350 | -1.855 |
| 51919.66451 | -1.837 |
| 51919.66554 | -1.855 |

(This table is available in its entirety in a machine-readable form in the online journal. A portion is shown here for guidance regarding its form and content.)

Table 5. NFO Differential Photometry of CF Tau

| HJD-2,400,000 | ΔV |
|---|---|
| 53359.62535 | -0.924 |
| 53359.62625 | -0.917 |
| 53359.62720 | -0.903 |
| 53359.62809 | -0.921 |
| 53359.62904 | -0.898 |

(This table is available in its entirety in a machine-readable form in the online journal. A portion is shown here for guidance regarding its form and content.)

Table 6. Photometric orbital parameters of CF Tau

| Parameter | URSA | NFO | Adopted |
|---|---|---|---|
| | | | |
| $J_{Aa}$ | 0.5206 ± 0.0012 | 0.5265 ± 0.0010 | 0.524 ± 0.003 |
| $r_{Aa} + r_{Ab}$ | 0.4303 ± 0.0010 | 0.4288 ± 0.0007 | 0.4296 ± 0.0008 |
| $r_{Aa}$ | 0.2477 ± 0.0006 | 0.2482 ± 0.0004 | 0.2480 ± 0.0007 |
| $R_{Ab}$ | 0.1826 ± 0.0011 | 0.1806 ± 0.0008 | 0.1816 ± 0.0013 |
| $k = r_{Aa}/r_{Ab}$ | 1.357 ± 0.010 | 1.375 ± 0.007 | 1.366 ± 0.009 |
| i (degrees) | 87.32 ± 0.20 | 87.21 ± 0.15 | 87.26 ± 0.25 |
| $u_{Aa}, u_{Ab}$ | 0.569, 0.411 fixed | 0.569, 0.411 fixed | 0.569, 0.411 fixed |
| $u_{Aa}', u_{Ab}'$ | 0.189, 0.288 fixed | 0.189, 0.288 fixed | 0.189, 0.288 fixed |
| $y_{Aa}, y_{Ab}$ | 0.42, 0.35 fixed | 0.42, 0.35 fixed | 0.42, 0.35 fixed |
| $q = M_{Aa}/M_{Ab}$ | 1.024 fixed | 1.024 fixed | 1.024 ± 0.005 |
| $L_{Aa}$ | 0.410 ± 0.005 | 0.425 ± 004 | 0.418 ± 0.008 |
| $L_{Ab}$ | 0.437 ± 0.005 | 0.436 ± 0.004 | 0.436 ± 0.004 |
| $L_B$ | 0.153 ± 0.006 | 0.139 ± 0.005 | 0.146 ± 0.009 |
| $L_{Aa}/L_{Ab}$ | 0.939 ± 0.015 | 0.975 ± 0.012 | 0.956 ± 0.023 |

| | | | | |
|---|---|---|---|---|
| σ (mmag) | 19.63038 | 12.01743 | | |
| N | 8052 | 5004 | | |
| Corrections | 83 | 105 | | |
| | | | | |

Note: the luminosity values cited here are fractional luminosities, where $L_{Aa}+L_{Ab}+L_B=1$.

Table 7. Comparison to the photometric orbit of Liakos et al. (2011) and refitted SWASP data

| Parameter in V-band | Liakos et al. (2011) | This Paper URSA + NFO | This Paper SWASP | Adopted SWASP |
|---|---|---|---|---|
| Inclination i (degrees) | 86.8 ± 0.1 | 87.26 ± 0.25 | 90 ± 2 | 87.26 fixed |
| Mass ratio q = $M_{Aa}/M_{Ab}$ | 0.940 ± 0.004 | 1.024 ± 0.005 | 1.024 ± 0.005 | 1.024 ± 0.005 |
| Temperature $T_{Aa}$ (K) | 5308 ± 5 | 5200 ± 150 | 5200 ± 150 | 5200 ± 150 |
| Temperature $T_{Ab}$ (K) | 5950 | 6000 ± 150 | 6000 ± 150 | 6000 ± 150 |
| Radius $r_{Aa}$ | 0.225 | 0.2480 ± 0.0007 | 0.2434 ± 0.0008 | 0.2465 ± 0.0008 |
| Radius $r_{Ab}$ | 0.206 | 0.1816 ± | 0.1842 ± 0.0015 | 0.1794 ± 0.0015 |

|  |  | 0.0013 |  |  |
| --- | --- | --- | --- | --- |
| Luminosity $L_{Aa}$ | 0.397 ± 0.001 | 0.418 ± 0.008 | 0.408 ± 0.012 | 0.446 ± 0.012 |
| Luminosity $L_{Ab}$ | 0.560 ± 0.002 | 0.436 ± 0.004 | 0.422 ± 0.008 | 0.423 ± 0.008 |
| Luminosity $L_B$ | 0.043 ± 0.002 | 0.146 ± 0.009 | 0.170 ± 0.004 | 0.131 ± 0.004 |
| σ (mmag) |  | 12.017 | 20.339 | 20.342 |
| N | 3521 | 13056 | 3521 | 3521 |

Note: the luminosity values cited here are fractional luminosities, where $L_{Aa}+L_{Ab}+L_B=1$.

Table 8. Physical parameters of the CF Tau eclipsing system

| Parameter | Star Aa | Star Ab |
| --- | --- | --- |
|  |  |  |
| Mass (solar masses) | 1.282 ± 0.009 | 1.251 ± 0.011 |
| Radius (solar radii) | 2.797 ± 0.011 | 2.048 ± 0.016 |
| Log g (cgs) | 3.653 ± 0.005 | 3.913 ± 0.008 |
| $T_{eff}$ (K) | 5200 ± 150 | 6000 ± 150 |
| Log L (solar units) | 0.71 ± 0.05 | 0.69 ± 0.05 |
| Measured v sin i (km/s) | 56 ± 4 | 38 ± 2 |
| Synchronous v sin i (km/s) | 51.3 ± 0.2 | 37.6 ± 0.3 |
| Semi-major axis (solar radii) | 11.28 ± 0.03 | |
| E(b-y) (mag) | 0.21 ± 0.05 | |
| $F_V$ | 3.695 ± 0.017 | 3.770 ± 0.014 |
| $M_V$ (mag) | 3.07 ± 0.17 | 3.00 ± 0.14 |
| m-M (mag) | 7.14 ± 0.24 | |
| Distance (pc) | 268 ± 30 | |
| $BC_V$ (mag) | -0.23 ± 0.11 | -0.04 ± 0.10 |
| $M_{bol}$ (mag) | 2.96 ± 0.12 | 3.01 ± 0.11 |

Notes: Absolute magnitudes $M_V$ are computed directly from the visual surface brightness $F_V$ following Popper (1980). $BC_V$ values are from Flower (1996), and their uncertainties include 0.1 mag added in quadrature to the uncertainty that comes from the temperature errors. Bolometric magnitudes assume $M_{bol}$(Sun) = 4.73 (see Torres 2010).